\documentclass[11pt,preprint,natbib]{aastex}

\shorttitle{Recoil of Black-Hole Binaries}
\shortauthors{Blanchet et al.}

\usepackage{bm} 
\usepackage{graphicx} 
\usepackage{epsfig}

\begin{document}


\title{Gravitational Recoil of Inspiralling Black-Hole Binaries \\to Second
  Post-Newtonian Order}



\author{Luc Blanchet} 
\affil{${\mathcal{G}}{\mathbb{R}}
  \varepsilon{\mathbb{C}}{\mathcal{O}}$ -- Gravitation et Cosmologie,\\
  Institut d'Astrophysique de Paris, C.N.R.S.,\\ 98 bis
  boulevard Arago, 75014 Paris, France; blanchet@iap.fr}

\author{Moh'd S. S. Qusailah\altaffilmark{1}} 
\affil{Raman Research Institute, Bangalore 560 080, 
India; mssq@rri.res.in}

\and

\author{Clifford M. Will\altaffilmark{2}} 
\affil{McDonnell Center for
the Space Sciences, Department of Physics, \\
Washington University, St.
Louis, Missouri 63130; cmw@wuphys.wustl.edu}

\altaffiltext{1}{On leave from University of Sana, Yemen}
\altaffiltext{2}{Visiting Researcher at ${\mathcal{G}}{\mathbb{R}}
\varepsilon{\mathbb{C}}{\mathcal{O}}$ -- Gravitation et Cosmologie,
Institut d'Astrophysique de Paris, C.N.R.S., 98 bis boulevard Arago,
75014 Paris, France}


\begin{abstract}
The loss of linear momentum by gravitational radiation and the resulting
gravitational recoil of black-hole binary systems may play an important
role in the growth of massive black holes in early galaxies. We
calculate the gravitational recoil of non-spinning black-hole binaries
at the second post-Newtonian order (2PN) beyond the dominant effect,
obtaining, for the first time, the 1.5PN correction term due to tails of
waves and the next 2PN term. We find that the maximum value of the net
recoil experienced by the binary due to the inspiral phase up to the
innermost stable circular orbit (ISCO) is of the order of $22 \,
\mathrm{km \, s}^{-1}$. We then estimate the kick velocity accumulated
during the plunge from the ISCO up to the horizon by integrating the
momentum flux using the 2PN formula along a plunge geodesic of the
Schwarzschild metric. We find that the contribution of the plunge
dominates over that of the inspiral. For a mass ratio $m_2/m_1=1/8$, 
we estimate a
total recoil velocity (due to both adiabatic and plunge phases) of $100
\pm 20 \, \mathrm{km \, s^{-1}}$. For a ratio 0.38, the recoil is
maximum and we estimate it to be $250 \pm 50 \, \mathrm{km \,
s^{-1}}$. In the limit of small mass ratio, we estimate $V/c \approx
0.043 \,(1 \pm 20\%)\, (m_2/m_1)^2$. Our estimates are
consistent with, but span a substantially narrower range than, those of
\cite{Favata04}.
\end{abstract}


\section{Introduction and summary}\label{secI}

The gravitational recoil of a system in response to the anisotropic
emission of gravitational waves is a phenomenon with potentially
important astrophysical consequences \citep{Merritt04}. Specifically, in
models for massive black hole formation involving successive mergers
from smaller black hole seeds, a recoil with a velocity sufficient to
eject the system from the host galaxy or mini-halo would effectively
terminate the process. Recoils could eject coalescing black holes from
dwarf galaxies or globular clusters. Even in galaxies whose potential
wells are deep enough to confine the recoiling system, displacement of
the system from the center could have important dynamical consequences
for the galactic core. Consequently, it is important to have a robust
estimate for the recoil velocity from inspiraling black hole binaries.

Recently, \cite{Favata04} estimated the kick
velocity for inspirals of both non-spinning and spinning black holes.
For example, for non-spinning holes, with a mass ratio of 1:8, they
estimated kick velocities between $20$ and $200 \, \mathrm{km \,
s^{-1}}$. The result was obtained by (i) making an estimate of the kick
velocity accumulated during the adiabatic inspiral of the system up to
its innermost stable circular orbit (ISCO), calculated using black-hole 
perturbation theory (valid in the small
mass ratio limit), extended to finite
mass ratios using scaling results from the quadrupole approximation,
and (ii) combining that with a crude estimate of the
kick velocity accumulated during the plunge phase (from the ISCO up to
the horizon). The plunge contribution generally dominates the recoil,
and is the most uncertain.

It is the purpose of this paper to compute more precisely the
gravitational recoil velocity during the inspiral phase up to the ISCO,
and to attempt to narrow that uncertainty in the plunge contribution for
non-spinning inspiralling black holes.

Earlier approaches for computing the recoil of general matter systems
include a near-zone computation of the recoil in linearized gravity
\citep{Peres62}, flux computations of the recoil as an interaction between
the quadrupole and octupole moments \citep{BoR61,Papa71}, a general
multipole expansion for the linear momentum flux \citep{Th80}, and a
radiation-reaction computation of the leading-order
post-Newtonian recoil \citep{B97}.

Using the post-Minkowskian and matching approach \citep{BD86,B95,B98mult}
for calculating equations of motion and gravitational radiation from
compact binary systems in a post-Newtonian (PN) sequence, 
\cite{BIJ02,BDEI04} have derived the gravitational energy
loss and phase to $\mathcal{O}[(v/c)^7]$ beyond the lowest-order
quadrupole approximation, corresponding to 3.5PN order, and the
gravitational wave amplitude to 2.5PN order~\citep{ABIQ04}. 
Using results from this
program, we derive the linear momentum flux from compact binary inspiral
to $\mathcal{O}[(v/c)^4]$, or 2PN order, beyond the lowest-order result.

The leading, ``Newtonian'' contribution\,\footnote{For want of a better
terminology, we denote the leading-order contribution to the recoil as
``Newtonian'', although it really corresponds to a 3.5PN radiation-reaction
effect in the
local equations of motion.} for binaries was first derived by 
\cite{Fitchett83}, and was extended to 1PN order by 
\cite{Wiseman92}. We extend these results by including both the 1.5PN
order contributions caused by gravitational-wave tail effects, and the
next 2PN order terms. We find that the linear momentum loss for binary
systems in circular orbits is given by\,\footnote{In most of this paper
we use units in which $G=1=c$. We generally do not indicate the
neglected PN remainder terms (higher than 2PN).}
\begin{eqnarray}\label{momfluxfinal}
\frac{d P^i}{dt} &=& \frac{464}{105}\,\frac{\delta m}{m}\,\eta^2\,
x^{11/2} \left[1+\left(-\frac{452}{87} -\frac{1139}{522}\eta\right)x +
\frac{309}{58}\,\pi\,x^{3/2} \right. \nonumber\\ &&\qquad \left.
+\left(-\frac{71345}{22968}+\frac{36761}{2088}\eta
+\frac{147101}{68904}\eta^2\right)x^2\right] {\hat \lambda}^i \,,
\end{eqnarray}
where $m=m_1+m_2$, $\delta m=m_1-m_2$, $\eta = m_1m_2/m^2$ (we have
$0<\eta \leq 1/4$, with $\eta=1/4$ for equal masses), and where
$x=(m\,\omega)^{2/3}$ is the PN parameter of the order of
$\mathcal{O}[(v/c)^2]$, where $\omega = d\phi/dt$ is the orbital angular
velocity. The quantity ${\hat \lambda}^i$ is a unit tangential vector
directed in the same sense as the orbital velocity ${\bf v}={\bf v}_1 -
{\bf v}_2$. The term at order $x^{3/2} = \mathcal{O}[(v/c)^3]$ comes
from gravitational-wave tails. Notice that, as expected for non-spinning
systems, the flux vanishes for equal-mass systems ($\delta m=0$ or
$\eta=1/4$).

\begin{figure}[t]
\begin{center}
\epsfig{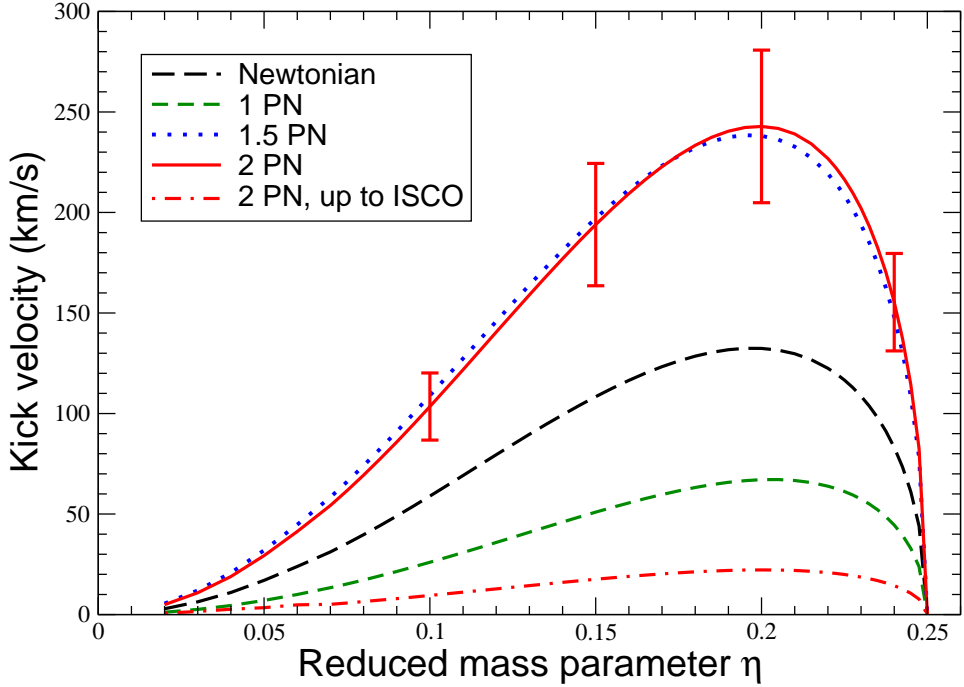}
\caption{Recoil velocity as a function of $\eta$.
\label{fig1}}
\end{center}
\end{figure}

To calculate the net recoil velocity, we integrate this flux along a
sequence of adiabatic quasi-circular inspiral orbits up to the ISCO. We
then connect that orbit to an unstable inspiral orbit of a test body
with mass $\mu = \eta\,m$ in the geometry of a Schwarzschild black hole
of mass $m$, with initial conditions that include the effects of
gravitational radiation damping. Using an integration variable that is
regular all the way to the event horizon of the black hole, we integrate
the momentum flux vector over the plunge orbit. Combining the adiabatic
and plunge contributions, calculating the magnitude, and dividing by $m$
gives the net recoil velocity. Figure \ref{fig1} shows the results.
Plotted as a function of the reduced mass parameter $\eta$ are curves
showing the results correct to Newtonian order, to 1PN order, to 1.5PN
order and to 2PN order. Also shown is the contribution of the adiabatic
part corresponding to the inspiral up to the ISCO (calculated to 2PN
order). The ``error bars'' shown are an attempt to assess the accuracy
of the result by including 2.5PN and 3PN terms with numerical
coefficients that are allowed to range over values between $-10$ and $10$.

We note that the 1PN result is smaller than the Newtonian result because
of the rather large negative coefficient seen in
Eq. (\ref{momfluxfinal}). On the other hand, the tail term at 1.5PN
order plays a crucial role in increasing the magnitude of the effect
(both for the adiabatic and plunge phases), and we observe that the
small 2PN coefficient in Eq. (\ref{momfluxfinal}) leads to the very
small difference between the 1.5PN and 2PN curves in Fig.~\ref{fig1}. In
our opinion this constitutes a good indication of the ``convergence'' of
the result. The momentum flux vanishes for the equal-mass case, $\eta =
1/4$, and reaches a maximum around $\eta = 0.2$ (a mass ratio of
$0.38$), which corresponds to the maximum of the overall factor $\eta^2
\delta m/m = \eta^2 \sqrt{1-4\eta}$, reflecting the relatively weak
dependence on $\eta$ in the PN corrections.  We propose in Eq.
(\ref{phenom}) below a phenomenological analytic formula which embodies this
weak $\eta$ dependence, and fits
our 2PN curve remarkably well. 

In contrast to the range 20 -- 200 $\mathrm{km \, s^{-1}}$ for
$\eta=0.1$ estimated by \cite{Favata04}, we estimate
a recoil velocity of $100 \pm 20 \, \mathrm{km \, s^{-1}}$ for this mass
ratio. For $\eta=0.2$ we estimate a recoil between $200$ and $300 \,
\mathrm{km \, s^{-1}}$, with a ``best guess'' of $250 \, \mathrm{km \,
s^{-1}}$ (the maximum velocity shown in Fig. \ref{fig1} is 
$243 \, \mathrm{km \, s^{-1}}$). 
We regard our computation of the recoil in the
adiabatic inspiral phase (up to the ISCO) as rather solid thanks to the
accurate 2PN formula we use, and the fact that the 1.5PN and 2PN results
are so close to each other. However, obviously, using PN methods to
study binary inspiral inside the ISCO is not without risks, and so it
would be very desirable to see a check of our estimates using either
black hole perturbation theory (along the lines of
\cite{OoNaka83}, \cite{NakaH83} or \cite{FitchettDet84}) or full numerical
relativity.  It is relevant to point out that our estimates agree well with
those obtained using numerical relativity in the ``Lazarus approach'', or
close-limit approximation, which treats the final merger of comparable-mass
black holes using a hybrid method combining
numerical relativity with perturbation theory~\citep{manuela05}.  In the
small mass-ratio limit, they also agree well 
with a calculation of the recoil from the head-on plunge from infinity using
perturbation theory~\citep{NakaOoKoj87}.
Therefore, we hope that our estimates will enable a more
focussed discussion of the astrophysical consequences of gravitational
radiation recoil.

The remainder of this paper provides details. In Section \ref{secII}, we
derive the 2PN accurate linear momentum flux using a multipole
decomposition, together with 2PN expressions for the multipole moments
in terms of source variables. In Section \ref{secIII} we specialize to
binary systems, and to circular orbits. In Section \ref{secIV}, we use
these results to estimate the recoil velocity and discuss various checks of
our estimates.  Section \ref{secV} makes
concluding remarks.

\section{General formulae for linear momentum flux}\label{secII}

The flux of linear momentum $\mathbf{P}$, carried away from general
isolated sources, is first expressed in terms of symmetric and
trace-free (STF) radiative multipole moments, which constitute very
convenient sets of observables parametrizing the asymptotic wave form at
the leading order $\vert\mathbf{X}\vert^{-1}$ in the distance to the
source, in an appropriate radiative coordinate system
$X^\mu=(T,\mathbf{X})$ \citep{Th80}. Denoting by $U_{i_1\cdots
i_\ell}(T)$ and $V_{i_1\cdots i_\ell}(T)$ the mass-type and current-type
radiative moments at radiative coordinate time $T$ (where $\ell$ is the
multipolar order), the linear momentum flux reads
\begin{eqnarray}\label{flux}
\mathcal{F}^i_\mathbf{P}(T) &=&
 \sum_{\ell=2}^{+\infty}\left\{\frac{2(\ell+2)(\ell+3)}{\ell
 (\ell+1)!(2\ell+3)!!}\,U_{ii_1\cdots i_\ell}^{(1)}(T)\,U_{i_1\cdots
 i_\ell}^{(1)}(T)\right.\nonumber\\&&\qquad\left.+\frac{8(\ell+2)}{(\ell-1)(\ell+1)!
 (2\ell+1)!!}\,\varepsilon_{ijk}\,U_{ji_1\cdots
 i_{\ell-1}}^{(1)}(T)\,V_{ki_1\cdots
 i_{\ell-1}}^{(1)}(T)\right.\nonumber\\&&\qquad\left.+\frac{8(\ell+3)}{(\ell+1)!
 (2\ell+3)!!}\,V_{ii_1\cdots i_\ell}^{(1)}(T)\,V_{i_1\cdots
 i_\ell}^{(1)}(T)\right\}\,,
\end{eqnarray}
where the superscript $(n)$ refers to the time-derivatives, and
$\varepsilon_{ijk}$ is Levi-Civita's antisymmetric symbol, such that
$\varepsilon_{123}=+1$. Taking into account all terms up to relative 2PN
order (in the case of slowly moving, PN sources), we obtain
\begin{eqnarray}\label{flux2PN}
\mathcal{F}^i_\mathbf{P} &=& \frac{2}{63}
\,U_{ijk}^{(1)}\,U_{jk}^{(1)}+\frac{16}{45} \,\varepsilon_{ijk}
\,U_{jl}^{(1)}\,V_{kl}^{(1)}\nonumber\\&+& \frac{1}{1134}
\,U_{ijkl}^{(1)}\,U_{jkl}^{(1)}+\frac{1}{126} \,\varepsilon_{ijk}
\,U_{jlm}^{(1)}\,V_{klm}^{(1)} + \frac{4}{63}
\,V_{ijk}^{(1)}\,V_{jk}^{(1)}\nonumber\\&+& \frac{1}{59400}
\,U_{ijklm}^{(1)}\,U_{jklm}^{(1)}+\frac{2}{14175} \,\varepsilon_{ijk}
\,U_{jlmn}^{(1)}\,V_{klmn}^{(1)} + \frac{2}{945}
\,V_{ijkl}^{(1)}\,V_{jkl}^{(1)}\,.
\end{eqnarray}
The first two terms represent the leading order in the linear momentum
flux, which corresponds to radiation reaction effects in the source's
equations of motion occuring at the 3.5PN order with respect to the
Newtonian force law. Indeed, recall that although the dominant radiation
reaction force is at 2.5PN order, the total integrated radiation
reaction force on the system 
(which gives the linear momentum loss or recoil) starts
only at the next 3.5PN order \citep{Peres62,BoR61,Papa71}. Radiation
reaction terms at the 3.5PN level for compact binaries in general orbits
have been computed by \cite{IW95}, \cite{JaraS97}, \cite{PW02}, \cite{KFS03}
and \cite{NB05}. In
Eq. (\ref{flux2PN}) all the terms up to 2PN order relative to the
leading linear momentum flux are included. This precision corresponds
formally to radiation reaction effects up to 5.5PN order.

The radiative multipole moments, seen at (Minkowskian) future null
infinity, $U_{i_1\cdots i_\ell}$ and $V_{i_1\cdots i_\ell}$, are now
related to the source multipole moments, say $I_{i_1\cdots i_\ell}$ and
$J_{i_1\cdots i_\ell}$, following the post-Minkowskian and matching
approach of \cite{BD86} and \cite{B95,B98mult}. The radiative moments differ
from the source moments by non-linear multipole interactions. At the
relative 2PN order considered in the present paper, the difference is
only due to interactions of the mass monopole $M$ of the source with
higher moments, so-called gravitational-wave tail effects. For the
source moments $I_{i_1\cdots i_\ell}$ and $J_{i_1\cdots i_\ell}$, we use
the expressions obtained in \cite{B95,B98mult}, valid for a
general extended isolated PN source. These moments are the analogues of
the multipole moments originally introduced by 
\cite{EW75} and generalized by \cite{Th80}, and which constitute
the building blocks of the direct integration of the retarded Einstein
equations (DIRE) formalism \citep{WWi96,PW00}. The radiative moments appearing
in Eq. (\ref{flux2PN}) are given in terms of the source moments by (see
Eqs.~(4.35) in \cite{B95})
\begin{mathletters}\label{tails}\begin{eqnarray}
U_{ij} (T) &=& I^{(2)}_{ij} (T) + 2\,M\int_{-\infty}^T
  d\tau\,I^{(4)}_{ij} (\tau)\left[ \ln \left(\frac{T-\tau}{2b}\right) +
  \frac{11}{12}\right]\,,\\ U_{ijk} (T) &=& I^{(3)}_{ijk} (T) +
  2\,M\int_{-\infty}^T d\tau\,I^{(5)}_{ijk} (\tau)\left[ \ln
  \left(\frac{T-\tau}{2b}\right) + \frac{97}{60}\right]\,,\\ V_{ij} (T)
  &=& J^{(2)}_{ij} (T) + 2\,M\int_{-\infty}^T d\tau\,J^{(4)}_{ij}
  (\tau)\left[ \ln \left(\frac{T-\tau}{2b}\right) +
  \frac{7}{6}\right]\,,
\end{eqnarray}\end{mathletters}
where $M\equiv I$ denotes the constant mass monopole or total ADM mass
of the source. The relative order of the tail integrals in
Eqs. (\ref{tails}) is 1.5PN. The constant $b$ entering the logarithmic
kernel of the tail integrals represents an arbitrary scale which is
defined by
\begin{equation}\label{b}
T = t_\mathrm{H} -\rho_\mathrm{H} -2\,M
\ln\left(\frac{\rho_\mathrm{H}}{b}\right)\,,
\end{equation}
where $t_\mathrm{H}$ and $\rho_\mathrm{H}$ correspond to a harmonic
coordinate chart covering the local isolated source ($\rho_\mathrm{H}$
is the distance of the source in harmonic coordinates). We insert
Eqs. (\ref{tails}) into the linear momentum flux (\ref{flux2PN}) and
naturally decompose it into
\begin{equation}\label{instail}
\mathcal{F}^i_\mathbf{P} =
\left(\mathcal{F}^i_\mathbf{P}\right)_\mathrm{inst} +
\left(\mathcal{F}^i_\mathbf{P}\right)_\mathrm{tail}\,,
\end{equation}
where the ``instantaneous'' piece, which depends on the state of the
source only at time $T$, is given by
\begin{eqnarray}\label{instpiece}
\left(\mathcal{F}^i_\mathbf{P}\right)_\mathrm{inst} &=& \frac{2}{63}
\,I_{ijk}^{(4)}\,I_{jk}^{(3)}+\frac{16}{45} \,\varepsilon_{ijk}
\,I_{jl}^{(3)}\,J_{kl}^{(3)}\nonumber\\&+& \frac{1}{1134}
\,I_{ijkl}^{(5)}\,I_{jkl}^{(4)}+\frac{1}{126} \,\varepsilon_{ijk}
\,I_{jlm}^{(4)}\,J_{klm}^{(4)} + \frac{4}{63}
\,J_{ijk}^{(4)}\,I_{jk}^{(3)}\nonumber\\&+& \frac{1}{59400}
\,I_{ijklm}^{(6)}\,I_{jklm}^{(5)}+\frac{2}{14175} \,\varepsilon_{ijk}
\,I_{jlmn}^{(5)}\,J_{klmn}^{(5)} + \frac{2}{945}
\,J_{ijkl}^{(5)}\,I_{jkl}^{(4)}\,,
\end{eqnarray}
and the ``tail'' piece, formally depending on the integrated past of the
source, reads
\begin{eqnarray}\label{tailpiece}
\left(\mathcal{F}^i_\mathbf{P}\right)_\mathrm{tail} &=& \frac{4\,M}{63}
\,I_{ijk}^{(4)}(T)\int_{-\infty}^T d\tau \,I_{jk}^{(5)}(\tau)
\left[\ln\left(\frac{T-\tau}{2b}\right)+\frac{11}{12}\right]
\nonumber\\&+&\frac{4\,M}{63} \,I_{jk}^{(3)}(T)\int_{-\infty}^T d\tau
\,I_{ijk}^{(6)}(\tau)
\left[\ln\left(\frac{T-\tau}{2b}\right)+\frac{97}{60}\right]
\nonumber\\&+&\frac{32\,M}{45}\,\varepsilon_{ijk}
\,I_{jl}^{(3)}(T)\int_{-\infty}^T d\tau \,J_{kl}^{(5)}(\tau)
\left[\ln\left(\frac{T-\tau}{2b}\right)+\frac{7}{6}\right]
\nonumber\\&+&\frac{32\,M}{45}\varepsilon_{ijk}
J_{kl}^{(3)}(T)\int_{-\infty}^T d\tau \,I_{jl}^{(5)}(\tau)
\left[\ln\left(\frac{T-\tau}{2b}\right)+\frac{11}{12}\right]\,.
\end{eqnarray}
The four terms in Eq. (\ref{tailpiece}) correspond to the tail parts of
the moments parametrizing the ``Newtonian'' approximation to the
flux given by the first line of (\ref{flux2PN}). All of them will
contribute at 1.5PN order.

\section{Application to compact binary systems}\label{secIII}

We specialize the expressions given in Section \ref{secII}, which are
valid for general PN sources, to the case of compact binary systems
modelled by two point masses $m_1$ and $m_2$. For this application, all
the required source multipole moments up to 2PN order admit known
explicit expressions, computed in \cite{BDI95,BIJ02} and \cite{ABIQ04} for
circular binary orbits. Here we quote only the results. Mass parameters
are $m=m_1+m_2$, $\delta m=m_1-m_2$ and the symmetric mass ratio $\eta =
m_1 m_2/m^2$. We define $\mathbf{x} \equiv \mathbf{x}_1-\mathbf{x}_2$
and $r \equiv \vert\mathbf{x}\vert$ to be the relative vector and
separation between the particles in harmonic coordinates, respectively,
and $\mathbf{v} \equiv d\mathbf{x}/dt$ to be their relative velocity
($t\equiv t_\mathrm{H}$ is the harmonic coordinate time). We have, for
mass-type moments,
\begin{mathletters}\label{massmom}\begin{eqnarray}
I_{ij} &=& \eta\,m \left\{{x}^{\langle ij\rangle}\left[1 +
\frac{m}{r}\left(-\frac{1}{42}-\frac{13}{14}\eta \right) +
\left(\frac{m}{r}\right)^2 \left(-\frac{461}{1512}
-\frac{18395}{1512}\eta - \frac{241}{1512} \eta^2\right)\right]
\right.\nonumber\\ &&~~\left. + \,r^2 v^{\langle
ij\rangle}\left[\frac{11}{21}-\frac{11}{7}\eta + \frac{m}{r}
\left(\frac{1607}{378}-\frac{1681}{378} \eta
+\frac{229}{378}\eta^2\right)\right]\right\}\,,\\ I_{ijk} &=&
-\eta\,\delta m\, \left\{ {x}^{\langle ijk\rangle} \left[1 - \frac{m}{r}
\eta- \left(\frac{m}{r}\right)^2
\left(\frac{139}{330}+\frac{11923}{660}\eta +\frac{29}{110}\eta^2\right)
\right]\right.\nonumber\\ &&~~+r^2\,x^{\langle i}v^{jk\rangle} \left[1 -
2\eta - \frac{m}{r} \left(-\frac{1066}{165}+\left.\frac{1433}{330}\eta
-\frac{21}{55} \eta^2\right) \right]\right\}\,,\\ I_{ijkl} &=& \eta
\,m\, \left\{ x^{\langle ijkl\rangle}\left[1 - 3\eta + \frac{m}{r}
\left(\frac{3}{110} - \frac{25}{22}\eta +
\frac{69}{22}\eta^2\right)\right]\right.\nonumber\\
&&~~+\left.\frac{78}{55}\,r^2\,v^{\langle ij} x^{kl\rangle} ( 1 - 5\eta
+ 5\eta^2 ) \right\}\,,\\ I_{ijklm}&=&-\eta\,\delta m\, x^{\langle
ijklm\rangle}\left(1-2\eta\right)\,,
\end{eqnarray}\end{mathletters}
and, for current-type moments,
\begin{mathletters}\label{currentmom}\begin{eqnarray}
J_{ij} &=& -\eta\,\delta m\left\{ \varepsilon^{ab\langle i} x^{j\rangle
a}v^b \left[1 +\frac{m}{r} \left(\frac{67}{28}-\frac{2}{7}\eta
\right)\right.\right.\nonumber \\ &&~~\left.\left.
+\left(\frac{m}{r}\right)^2\left(\frac{13}{9} -\frac{4651}{252}\eta
-\frac{1}{168}\eta^2 \right)\right]\right\}\,, \\ J_{ijk} &=& \eta\,m
\left\{ \varepsilon^{ab\langle i} x^{jk\rangle a} v^b \left[1 - 3\eta +
\frac{m}{r} \left(\frac{181}{90} - \frac{109}{18}\eta +
\frac{13}{18}\eta^2\right)\right] \right.\nonumber \\
&&~~+\left.\frac{7}{45}\,r^2 \,\varepsilon^{ab\langle i}v^{jk\rangle
b}x^a (1 - 5\eta + 5\eta^2)\right\}\,,\\ J_{ijkl} &=& -\eta\,\delta m
\,\varepsilon^{ab\langle i}x^{jkl\rangle a} v^b \left(1-2\eta\right)\,.
\end{eqnarray}\end{mathletters}
We indicate the symmetric-trace-free projection using carets surrounding
indices. Thus, the STF product of $\ell$ spatial vectors, say
$x^{i_1\cdots i_\ell}=x^{i_1}\cdots x^{i_\ell}$, is denoted $x^{\langle
i_1\cdots i_\ell\rangle}=\mathrm{STF}\left[ x^{i_1\cdots
i_\ell}\right]$. Similarly, we pose $x^{\langle i_1\cdots
i_k}v^{i_{k+1}\cdots i_\ell\rangle}=\mathrm{STF}\left[ x^{i_1\cdots
i_k}v^{i_{k+1}\cdots i_\ell}\right]$.

The total mass $M$ in front of the tail integrals in (\ref{tails})
simply reduces, at the approximation considered in this paper, to the
sum of the masses, {\em i.e.} $M=m=m_1+m_2$.  Thus, to compute the tail
contributions (\ref{tailpiece}), we simply need the Newtonian approximation
for all the moments.

As seen in Eqs. (\ref{instpiece})-- (\ref{tailpiece}) we need to perform
repeated time-differentiations of the moments. These are consistently
computed using for the replacement of accelerations the binary's 2PN
equations of motion in harmonic coordinates (for circular 2PN orbits)
\begin{equation}\label{EOM}
\frac{d v^i}{dt} = -\omega^2 x^i\,,
\end{equation}
where $\omega$ denotes the angular frequency of the circular motion,
which is related to the orbital separation $r$ by the generalized Kepler
law
\begin{equation}\label{angularvel}
\omega^2 = \frac{m}{r^3} \biggl\{ 1+\frac{m}{r}(-3+\eta) +
\left(\frac{m}{r}\right)^2\left( 6 +\frac{41}{4} \eta +\eta^2 \right)
\biggr\}\,.
\end{equation}
The inverse of this law yields [using $x\equiv (m\,\omega)^{2/3}$]
\begin{equation}\label{angularvelinv}
\frac{m}{r} = x\,\biggl\{ 1+x\,\left(1-\frac{\eta}{3}\right) +
x^2\,\left(1 - \frac{65}{12} \eta\right) \biggr\}\,.
\end{equation}

The tail integrals of Eq. (\ref{tailpiece}) are computed in the
adiabatic approximation by substituting into the integrands the
components of the moments calculated for exactly circular orbits, with
the current value of the orbital frequency $\omega$ (at time $T$), but
with different phases corresponding to whether the moment is evaluated
at the current time $T$ or at the retarded time $\tau <T$. For exactly
circular orbits the phase difference is simply
$\Delta\phi=\omega(T-\tau)$. All the contractions of indices are
performed, and the result is obtained in the form of a sum of terms which
can all be analytically computed by means of the mathematical formula
\begin{equation}\label{formula}
\int_0^{+\infty} d\tau\,\ln\left(\frac{\tau}{2
b}\right)\,e^{\mathrm{i}\,n\,\omega\,\tau} =
-\frac{1}{n\,\omega}\left\{\frac{\pi}{2} +\mathrm{i}
\Bigl[\ln\left(2n\,\omega \,b\right)+C\Bigr]\right\}\,,
\end{equation}
where $\omega$ is the orbital frequency, $n$ the number of the
considered harmonics of the signal ($n=1$, $2$ or $3$ at the present 2PN
order) and $C=0.577\cdots$ is Euler's constant. As shown in
\cite{BS93} (see also \cite{BDI95,ABIQ04}), this procedure to
compute the tails is correct in the adiabatic limit, {\em i.e.} modulo
the neglect of 2.5PN radiation reaction terms $\mathcal{O}[(v/c)^5]$ which do
not contribute at the present order.

As it will turn out, the effect of tails in the linear momentum flux
comes only from the first term in the right side of
Eq. (\ref{formula}), proportional to $\pi$. All the contributions due
to the second term in (\ref{formula}), which involves the logarithm of
frequency, can be reabsorbed into a convenient definition of the phase
variable, and then shown to correspond to a very small phase
modulation which is negligible at the present PN order. This
possibility of introducing a new phase variable containing all the
logarithms of frequency was usefully applied in previous computations
of the binary's polarization waveforms \citep{BIWW96,ABIQ04}. We
introduce the phase variable $\psi$ differing from the actual orbital
phase angle $\phi$, whose time derivative equals the orbital frequency
($\dot{\phi}=\omega$), by
\begin{equation}\label{psi}
\psi = \phi -
2m\,\omega\ln\left(\frac{\omega}{\widehat{\omega}}\right)\,,
\end{equation}
where $\widehat{\omega}$ denotes a certain constant frequency scale that
is related to the constant $b$ which was introduced into the tail
integrals (\ref{tails}), and parametrizes the coordinate transformation
(\ref{b}) between harmonic and radiative coordinates. The constants
$\widehat{\omega}$ and $b$ are in fact devoid of any physical meaning
and can be chosen at will \citep{BIWW96,ABIQ04}. To check this let us use
the time dependence of the orbital phase $\phi$ due to
radiation-reaction inspiral in the adiabatic limit, given at the lowest
quadrupolar order by (see {\em e.g.} \cite{BIWW96})
\begin{equation}\label{radreac}
\phi_c-\phi(T)=\frac{1}{\eta}\left(\frac{\eta}{5m}[T_c-T]\right)^{5/8}\,,
\end{equation}
where $T_c$ and $\phi_c$ denote the instant of coalescence and the value
of the phase at that instant. Then it is easy to verify that an
arbitrary rescaling of the constant $\widehat{\omega}$ by
$\widehat{\omega}\rightarrow\lambda\,\widehat{\omega}$ simply
corresponds to a constant shift in the value of the instant of
coalescence, namely $T_c\rightarrow T_c+2m\,\ln\lambda$. Thus, any
choice for $\widehat{\omega}$ is in fact irrelevant since it is
equivalent to a choice of the origin of time in the wave zone. The
relation between $\widehat{\omega}$ and $b$ is given here for
completeness,
\begin{equation}\label{omegahat}
\widehat{\omega} =
\frac{1}{b}\,\mathrm{exp}\!\left[\frac{5921}{1740}+\frac{48}{29}\ln
2-\frac{405}{116}\ln 3 - C\right]\,. 
\end{equation}
The irrelevance of $\widehat{\omega}$ and $b$ is also clear from
Eq. (\ref{b}) where one sees that they correspond to an adjustement of
the time origin of radiative coordinates with respect to that of the
source-rooted harmonic coordinates.

Let us next point out that the phase modulation of the log-term in
Eq. (\ref{psi}) represents in fact a very small effect, which is formally of
order 4PN relative to the dominant radiation-reaction expression of
the phase as a function of time, given by (\ref{radreac}). This is clear
from the fact that Eq. (\ref{radreac}) is of the order of the inverse of
radiation-reaction effects, which can be said to correspond to $-$2.5PN
order, and that, in comparison, the tail term is of order +1.5PN, which
means 4PN relative order. In the
present paper we shall neglect such 4PN effects and will therefore
identify the phase $\psi$ with the actual orbital phase of the binary.

We introduce two unit vectors $\hat{n}^i$ and $\hat{\lambda}^i$,
respectively along the binary's separation, \textit{i.e.} in the
direction of the phase angle $\psi$, and along the relative velocity, in
the direction of $\psi+\frac{\pi}{2}$, namely
\begin{equation}\label{unitvect}
\hat{n}^i = \left(\begin{array}{c}\cos \psi\\\sin
\psi\\0\end{array}\right)~~\,\rm{and}\,~~\hat{\lambda}^i =
\left(\begin{array}{c}-\sin \psi\\\cos \psi\\0\end{array}\right).
\end{equation}
Finally, the reduction of the two terms (\ref{instpiece}) and
(\ref{tailpiece}) for compact binaries using the source moments
(\ref{massmom})--(\ref{currentmom}) is straightforward, and
yields the complete expression of
the 2PN linear momentum flux, 
\begin{eqnarray}\label{momflux}
\mathcal{F}^i_\mathbf{P} &=& -\frac{464}{105}\,\eta^2\,\frac{\delta
  m}{m}\left(\frac{m}{r}\right)^{11/2}\left[1+\left(-\frac{1861}{174}
  -\frac{91}{261}\eta\right)\frac{m}{r}
  +\frac{309}{58}\,\pi\left(\frac{m}{r}\right)^{3/2}\right.\nonumber\\
  &&\qquad\left. +\left(\frac{139355}{2871}+\frac{36269}{1044}\eta
  +\frac{17}{3828}\eta^2\right)\left(\frac{m}{r}\right)^2\right]\,
  \hat{\lambda}^i \,.
\end{eqnarray}
The first term is the ``Newtonian'' one which, as we noted above, really
corresponds to a 3.5PN radiation reaction effect. It is followed by the
1PN relative correction, then the 1.5PN correction, proportional to $\pi$ and
which is exclusively due to tails, and finally the 2PN correction term.
We find that the 1PN term is in agreement with the previous result by
\cite{Wiseman92}. The tail term at order 1.5PN and the 2PN term
are new with the present paper. Alternatively we can also express the
flux in terms of the orbital frequency $\omega$, with the help of the PN
parameter defined by $x = (m\,\omega)^{2/3}$. Using
Eq. (\ref{angularvelinv}) we obtain
\begin{eqnarray}\label{momfluxx}
\mathcal{F}^i_\mathbf{P} &=& -\frac{464}{105}\,\frac{\delta
m}{m}\,\eta^2\, x^{11/2} \left[1+\left(-\frac{452}{87}
-\frac{1139}{522}\eta\right)x + \frac{309}{58}\,\pi\,x^{3/2} \right.
\nonumber\\ &&\qquad \left.
+\left(-\frac{71345}{22968}+\frac{36761}{2088}\eta
+\frac{147101}{68904}\eta^2\right)x^2\right] {\hat \lambda}^i \,.
\end{eqnarray}
The latter form is interesting because it remains invariant under a
large class of gauge transformations.

Next, in order to obtain the local loss of linear momentum by the
source, we apply the momentum balance equation
\begin{equation}\label{balance}
\frac{d P^i}{d T} = - \mathcal{F}^i_\mathbf{P} (T)\,,
\end{equation}
which yields Eq. (\ref{momfluxfinal}). Upon integration, this yields
the net change of linear momentum, say $\Delta P^i= - \int_{-\infty}^T
dt\,\mathcal{F}^i_\mathbf{P}(t)$. In the adiabatic limit,
\textit{i.e.} at any instant before the passage at the ISCO, the
closed form of $\Delta P^i$ can be simply obtained (for circular
orbits) from the fact that ${d\hat{n}^i}/{dt}=\omega \hat{\lambda}^i$
and the constancy of the orbital frequency $\omega$.  This is of
course correct modulo fractional error terms $\mathcal{O}[(v/c)^5]$
which are negligible here. So, integrating the balance equation
(\ref{balance}) in the adiabatic approximation simply amounts to
replacing the unit vector $\hat{\lambda}^i$ by $\hat{n}^i$ and
dividing by the orbital frequency $\omega$. In this way we obtain the
recoil velocity $V^i\equiv\Delta P^i/m$ as\,\footnote{The recoil could
also be defined from the special-relativistic relation $V^i = \Delta
P^i/\sqrt{m^2+\Delta\mathbf{P}^2}$, but since $\Delta P^i$ is of order
3.5PN the latter ``relativistic'' definition yields the same 2PN
results, and in fact differs from our own definition by extremely
small corrections, at the 7PN order.}
\begin{eqnarray}\label{recoil}
V^i &=& \frac{464}{105}\,\eta^2\,\frac{\delta
  m}{m}\left(\frac{m}{r}\right)^4\left[1+\left(-\frac{800}{87}
  -\frac{443}{522}\eta\right)\frac{m}{r}
  +\frac{309}{58}\,\pi\left(\frac{m}{r}\right)^{3/2}\right.\nonumber\\
  &&\qquad\left. +\left(\frac{754975}{22968}+\frac{67213}{2088}\eta
  +\frac{1235}{22968}\eta^2\right)\left(\frac{m}{r}\right)^2\right]\,
  \hat{n}^i \,,
\end{eqnarray}
or, alternatively, in terms of the $x$-parameter,
\begin{eqnarray}\label{recoilx}
V^i &=& \frac{464}{105}\,\eta^2\,\frac{\delta
  m}{m}x^4\left[1+\left(-\frac{452}{87} -\frac{1139}{522}\eta\right)\,x
  +\frac{309}{58}\,\pi\,x^{3/2}\right.\nonumber\\ &&\qquad\left.
  +\left(-\frac{71345}{22968}+\frac{36761}{2088}\eta
  +\frac{147101}{68904}\eta^2\right)\,x^2\right]\, \hat{n}^i \,.
\end{eqnarray}
Equations (\ref{momfluxfinal}) and (\ref{recoilx})
will be the basis for our numerical estimates of
the recoil velocity, to be carried out in the next Section.

\section{Estimating the recoil velocity}\label{secIV}

\subsection{Basic assumptions and analytic formulae}
\label{secIVA}

We now wish to use Eqs. (\ref{momfluxfinal}) and 
(\ref{recoilx}) to estimate the recoil velocity
that results from the inspiral and merger of two black holes. It is
clear that the PN approximation becomes less reliable inside the
innermost stable circular orbit (ISCO). Nevertheless, we have an
expression that is accurate to 2PN order beyond the leading effect,
which will therefore be very accurate over all the inspiral phase all
the way down to the ISCO, so we have some hope that, if the higher-order
terms can be seen to be small corrections throughout the process, we can
make a robust estimate of the overall kick.

In Eq. (\ref{recoilx}) we have re-expressed the recoil velocity in
terms of the orbital angular velocity $\omega$,
Eq. (\ref{angularvel}), consistently to 2PN order. One advantage of
this change of variables is that the momentum loss is now expressed in
terms of a somewhat less coordinate dependent quantity, namely the
orbital angular velocity as seen from infinity. A second advantage is
that the convergence of the PN series is significantly improved. In
terms of the variable $m/r$, the coefficients of the 1PN and 2PN terms
are of order $-$10 and 33 -- 41, respectively, depending on the value
of $\eta$, whereas in terms of $x$, they are of order $-$5 and $-$3 --
$+$1.4, respectively.

We assume that the system undergoes an adiabatic inspiral along a
sequence of circular orbits up to the ISCO. For the present discussion
the ISCO is taken to be the one for point-mass motion around a
Schwarzschild black hole of mass $m=m_1+m_2$, namely
$m\,\omega_\mathrm{ISCO} = 6^{-3/2}$ or $x_\mathrm{ISCO}=1/6$. The
recoil velocity at the ISCO is thus given by
\begin{eqnarray}\label{kickisco}
V_\mathrm{ISCO}^i &=& \frac{464}{105}\,\frac{\delta
m}{m}\,\eta^2\,x_\mathrm{ISCO}^4 \left[1+\left(-\frac{452}{87}
-\frac{1139}{522}\eta\right)x_\mathrm{ISCO} +
\frac{309}{58}\,\pi\,x_\mathrm{ISCO}^{3/2}\right.\nonumber\\
&&\qquad\left. +\left(-\frac{71345}{22968}+\frac{36761}{2088}\eta
+\frac{147101}{68904}\eta^2\right)x_\mathrm{ISCO}^2\right] 
\hat{n}^i_\mathrm{ISCO} \,.
\end{eqnarray}

In order to determine the kick velocity accumulated during the plunge,
we make a number of simplifying assumptions. We first assume that the
plunge can be viewed as that of a ``test'' particle of mass $\mu$ moving
in the fixed Schwarzschild geometry of a body of mass $m$, following the
``effective one-body'' approach of 
\cite{BuonD98} and \cite{damour01}. We also assume that the effect on the
plunge orbit of the radiation of energy and angular momentum may be
ignored; over the small number of orbits that make up the plunge, this
seems like a reasonable approximation 
(\cite{Favata04} make the same assumption).

We therefore adopt the geodesic equations for
the Schwarzschild geometry,
\begin{mathletters}\label{schwgeodesic}
\begin{eqnarray}
\frac{dt}{d\tau} &=& \frac{\tilde{E}}{1-{2m}/{r_\mathrm{S}}} \,, 
\label{schwgeodesica}
\\
\frac{d\psi}{d\tau} &=& \frac{\tilde{L}}{r_\mathrm{S}^2} \,, 
\label{schwgeodesicb}
\\
\left(\frac{dr_\mathrm{S}}{d\tau}\right)^2 &=& \tilde{E}^2 -
\left(1-\frac{2m}{r_\mathrm{S}}\right)\left[1+
\frac{\tilde{L}^2}{r_\mathrm{S}^2}\right] \,,
\label{schwgeodesicc}
\end{eqnarray}\end{mathletters}
where $\tau$ is proper time along the geodesic, $\tilde{E}$ is the energy per
unit mass ($\mu$ in this case), and $\tilde{L} \equiv m \, \bar{L}$ is the
angular momentum per unit mass. 
Then, from Eqs. (\ref{schwgeodesicb}) and (\ref{schwgeodesicc}), we
obtain the phase angle of the orbit $\psi$ as a function of
$y=m/r_\mathrm{S}$ by
\begin{equation}
\psi = \int_{y_0}^y \left ( \frac{\bar{L}^2}{\tilde{E}^2
-(1-2y)(1+\bar{L}^2 y^2)} \right )^{1/2} dy  \,,
\label{phase}
\end{equation}
where we choose $\psi=0$ at the beginning of the plunge orbit defined by
$y=y_0$. 

The kick velocity accumulated during the plunge is then given
by\footnote{The radiative time $T$ in the linear momentum loss law
(\ref{balance}) can be viewed as a dummy variable, and we henceforth
replace it by the Schwarzschild coordinate time $t$.}
\begin{equation}
\Delta V^i_\mathrm{plunge} = \frac{1}{m}
\int_{t_0}^{t_\mathrm{Horizon}} \frac{dP^i}{dt} dt \,.
\end{equation}
However, the coordinate time $t$ is singular at the event horizon, so we
must find a non-singular variable to carry out the integration. We
choose the ``proper'' angular frequency, $\bar{\omega} = d\psi/d\tau$.
In addition to being monotonically increasing, 
this variable has the following useful properties along the plunge
geodesic:
\begin{mathletters}\label{ombar}\begin{eqnarray}
m \,\bar{\omega} &=& \bar{L}\, y^2 \,,
\label{ombara}\\
m \,\omega &=& m \,\bar{\omega} \,\frac{1-2y}{\tilde{E}} 
= \frac{\bar{L}}{\tilde{E}}\,y^2(1-2y)\,, 
\label{ombarb}\\
\frac{d\bar{\omega}}{dt} &=& \frac{2}{m} \,\omega \,y \Bigl[\tilde{E}^2 -
(1-2y)(1+\bar{L}^2 y^2) \Bigr]^{1/2} \,.
\label{ombarc}
\end{eqnarray}\end{mathletters}
Then
\begin{eqnarray}\label{kickplunge}
\Delta V^i_\mathrm{plunge}&=& \frac{1}{m} \int \frac{dP^i}{dt}
\frac{d\bar{\omega}}{d\bar{\omega}/dt} \nonumber\\ &=& \bar{L}
\,\int_{y_0}^{y_\mathrm{Horizon}} \left ( \frac{1}{m\,\omega}
\frac{dP^i}{dt} \right ) \frac{dy}{\Bigl[\tilde{E}^2 -
(1-2y)(1+\bar{L}^2 y^2) \Bigr]^{1/2}} \,,
\end{eqnarray}
where $y_0$ is defined by the matching to a circular orbit at the ISCO
that we shall discuss below.

Notice that, because $dP^i/dt \propto x^{11/2} \propto
(m\,\omega)^{11/3}$, the quantity in parentheses in
Eq. (\ref{kickplunge}) is well behaved at the horizon; in fact it
vanishes at the horizon because $\omega=0$ there [cf.
Eq. (\ref{ombarb})]. 
Thus, we find that  the integrand of Eq. (\ref{kickplunge}) behaves like
$(m\,\omega)^{8/3} \propto (1-2y)^{8/3}$ at the horizon, and the integral is
perfectly convergent.
Furthermore, since the expansion of $dP^i/dt$ is in
powers of $m\,\omega$, the convergence of the PN series is actually
improved as the particle approaches the horizon. To carry out the
integral, then, we substitute for $x=(m\,\omega)^{2/3}$ in $dP^i/dt$
using Eq. (\ref{ombarb}), and integrate over $y$.

We regard this approach as robust, because it uses
invariant quantities such as angular frequencies, and uses the nature of
the flux formula itself to obtain an 
integral that is automatically convergent. 
\cite{Favata04} tried to control the singular
behavior of the $t$ integration with an {\em ad hoc} regularization scheme.

We then combine Eqs. (\ref{kickisco}) and (\ref{kickplunge}) vectorially
to obtain the net kick velocity,
\begin{equation}
\Delta V^i = V^i_\mathrm{ISCO} + \Delta V^i_\mathrm{plunge} \,,
\label{kicktotal}
\end{equation}
in which $V^i_\mathrm{ISCO}$ is given by Eq. (\ref{kickisco}) above
with $\hat{n}^i_\mathrm{ISCO} = (1,0,0)$.

There are many ways to match a circular orbit at the ISCO to a
suitable plunge orbit; we use two different methods. In one, we give
the particle an energy $\tilde{E}$ such that, at the ISCO, and for an
ISCO angular momentum $\tilde{L}_\mathrm{ISCO} = \sqrt{12}\,m$, the
particle has a radial velocity given by the standard quadrupole
energy-loss formula for a circular orbit, namely $dr_\mathrm{H}/dt =
-(64/5)\eta (m/r_\mathrm{H})^3$, where $r_\mathrm{H}$ is the orbital
separation in harmonic coordinates. At the ISCO for a test body,
$r_\mathrm{H}= 5m$, so we have $(dr_\mathrm{H}/dt)_\mathrm{ISCO}=
-(8/25)^2 \eta$. This means also $(dr_\mathrm{S}/dt)_\mathrm{ISCO}=
-(8/25)^2 \eta$ in the Schwarzschild coordinate
$r_\mathrm{S}=r_\mathrm{H}+m$ (recall that
$t_\mathrm{S}=t_\mathrm{H}=t$).  It is straightforward to show that
the required energy for such an orbit is given by
\begin{equation}
\tilde{E}^2 = \frac{8}{9} \left [ 1 - \frac{9}{4} \left (
\frac{dr_\mathrm{S}}{dt} \right )^2_\mathrm{ISCO} \right ]^{-1} \,.
\label{Eisco}
\end{equation}
We therefore integrate Eq. (\ref{kickplunge}) with that energy,
together with $\bar{L}_\mathrm{ISCO} = \sqrt{12}$ and the initial
condition $y_0=y_\mathrm{ISCO}=1/6$ (from Eq. (\ref{ombarb}) we note
that, with this choice of initial condition, $m\,\omega_0 \ne
6^{-3/2}$). We choose also to terminate the integration when
$r_\mathrm{S}=2(m+\mu)$ hence $y_\mathrm{Horizon}^{-1}=2(1+\eta)$.

With this initial condition, the number of orbits ranges from 1.2 for
$\eta=1/4$ to 1.8 for $\eta=1/10$ to 4.3 for $\eta=1/100$. It is also
useful to note that the radial velocity remains small compared to the
tangential velocity throughout most of the plunge; the ratio
$(dr_\mathrm{S}/d\tau)/(r_\mathrm{S}\,d\psi/d\tau)
=r_\mathrm{S}^{-1}dr_\mathrm{S}/d\psi$ reaches 0.14 at
$r_\mathrm{S}=4m$, 0.3 at $r_\mathrm{S}=3m$, and 0.5 at
$r_\mathrm{S}=2m$, roughly independently of the value of $\eta$. This
justifies our use of circular orbit formulae for the momentum flux as a
reasonable approximation.

In a second method, we evolve an orbit at the ISCO piecewise to a new
orbit inside the ISCO, as follows: using the energy and angular momentum
balance equations for circular orbits in the adiabatic limit at the
ISCO, we have
\begin{mathletters}\label{dEJdt}
\begin{eqnarray}
\frac{d\tilde{E}}{d t} &=& -\frac{32}{5}\,\frac{\eta}{m}
\,x_\mathrm{ISCO}^5\,,\\ 
\frac{d\tilde{L}}{d t} &=&
\omega_\mathrm{ISCO}^{-1} \,\frac{d\tilde{E}}{d t} = -\frac{32}{5}\,\eta
\,x_\mathrm{ISCO}^{7/2}\,.
\end{eqnarray}
\end{mathletters}
We approximate these relations by ``discretizing'' the variations of the
energy and angular momentum in the left sides around the ISCO values
$\tilde{E}_\mathrm{ISCO}=\sqrt{8/9}$ and
$\tilde{L}_\mathrm{ISCO}=\sqrt{12}\,m$. Hence, we write $d \tilde{E}/d
t=(\tilde{E}-\tilde{E}_\mathrm{ISCO})/(\alpha P)$ and $d \tilde{L}/d
t=(\tilde{L}-\tilde{L}_\mathrm{ISCO})/(\alpha P)$, where $\alpha P$ 
denotes a fraction of the orbital
period $P$ of the circular motion at the ISCO. Using then
$\omega_\mathrm{ISCO}=2\pi/P=m^{-1}x_\mathrm{ISCO}^{3/2}$ this gives the
following values for the plunge orbit
\begin{mathletters}\label{EJtilde}
\begin{eqnarray}
\tilde{E} &=& \tilde{E}_\mathrm{ISCO} -
\frac{64\pi}{5}\,\alpha \,\eta\,x_\mathrm{ISCO}^{7/2}\,,\\ 
\bar{L} &=&
\bar{L}_\mathrm{ISCO} -
\frac{64\pi}{5}\,\alpha \,\eta\,x_\mathrm{ISCO}^{2}\,.
\end{eqnarray}
\end{mathletters}
Then, in this second model we integrate Eq. (\ref{kickplunge}) with the
latter values, and using the initial inverse radius $y_0 =
(m/r_\mathrm{S})_\mathrm{initial}$ of this new orbit which is given by
the solution of the equation
\begin{equation}\label{y0}
m \,\omega_\mathrm{ISCO} = 6^{-3/2} = \frac{\bar{L}}{\tilde{E}} \,y_0^2
(1-2y_0) \,.
\end{equation}
For the final value we simply take the horizon at $r_\mathrm{S}=2m$
(hence $y_\mathrm{Horizon}=1/2$), in the spirit of the effective
one-body approach \citep{BuonD98,damour01} where the binary's total mass
$m$ is identified with the black-hole mass and where $\mu$ is the test
particle's mass.
For the fraction $\alpha$ of the period, we choose values between 1 and
0.01, and check the dependence of the result on this choice (see below).

\subsection{Numerical results and checks}\label{secIVB}

First, we display the recoil velocities at the ISCO given by
Eq. (\ref{kickisco}) for each PN order and various values of $\eta$ in
Table \ref{table1}. The 2PN values of the velocity at the ISCO are also
plotted as a function of $\eta$ in Figure \ref{fig1} (dot-dash curve).
On should note, from Table \ref{table1}, the somewhat strange behavior
of the 1PN order, which nearly cancels out the Newtonian approximation
(as already pointed out by \cite{Wiseman92}). The maximum velocity
accumulated in the inspiral phase is around $22\, \mathrm{km \,
s}^{-1}$.

\begin{table}[t]
\begin{center}
\caption{Recoil velocity ($\mathrm{km \, s}^{-1}$) at the ISCO defined
  by $x_\mathrm{ISCO}=1/6$.}
\label{table1}
\begin{tabular}{l||r|r|r|r|r}
$\eta={\mu}/{m}$&0.05&0.1&0.15&0.2&0.24\\ \hline\hline 
Newtonian &2.29&7.92&14.56&18.30&11.78\\ \hline 
N $+$ 1PN &0.27&0.77&1.16&1.12&0.55\\ \hline 
N $+$ 1PN $+$ 1.5PN (tail)&2.87&9.80&17.74&21.96&13.97\\ \hline 
N $+$ 1PN $+$ 1.5PN $+$ 2PN &2.73&9.51&17.57&22.22&14.38\\
\end{tabular}
\end{center}
\end{table}

Next, we evaluate the kick velocity from the plunge phase, and carry out
a number of tests of the result. In our first model, where the plunge
energy is given by (\ref{Eisco}), we choose $r_\mathrm{S}=6\,m$ as the
ISCO, and $r_\mathrm{S}=2(m+\mu) = 2\,m(1+\eta)$ as the final merger
point. The latter value corresponds to the sum of the event horizons of
black holes of mass $m$ and $\mu$, and is an effort to estimate the end
of the merger when a common event horizon envelops the two black holes,
and any momentum radiation shuts off. 

The resulting total kick velocity as a function of $\eta$ is plotted as
the solid (red) curve in Figure \ref{fig1}. We also consider the kick
velocity generated when we take only the leading ``Newtonian''
contribution (dashed [black] curve), and when we include the 1PN terms
(short dashed [green] curve) and the 1PN $+$ 1.5PN terms (dotted [blue]
curve). Notice that, because the 1PN term has a negative coefficient,
the net kick velocity at 1PN order is smaller than at Newtonian order.
On the other hand, because the 2PN coefficient is so small, the 1.5PN
correct value and the 2PN correct value are very close to each other.

In order to test the sensitivity of the result to the PN expansion, we
have considered terms of 2.5PN, 3PN and 3.5PN order, by adding to the
expression (\ref{momfluxfinal}) terms of the form $a_\mathrm{2.5PN}
x^{5/2} + a_\mathrm{3PN} x^{3} +a_\mathrm{3.5PN} x^{7/2}$, and varying
each coefficient between +10 and $-$10. For example, varying
$a_\mathrm{2.5PN}$ and $a_\mathrm{3PN}$, leads to a maximum variation
in the velocity of $\pm \,30 \%$ [{\em i.e.} between the values
($-$10,$-$10) and (10,10)] for a range of $\eta$. Assuming that the
probability of occurrence of a specific value of each coefficient is
uniform within the interval [$-$10,10], we estimate an rms error in
the kick velocity, shown as ``error bars'' in Figure
\ref{fig1}. Varying $a_\mathrm{3.5PN}$ between $-$10 and 10 has only a
10\% effect on the final velocity.  These considerations lead us to
crudely estimate that our results are probably good to $\pm 20 \%$.

In the limit of small $\eta$, our numerical results give an estimate for
the kick velocity:
\begin{equation}\label{pertlimit}
\frac{V}{c} \approx \,0.043 \,\eta^2 \sqrt{1-4\eta}~~\,
\rm{when}\,~~\eta\rightarrow 0\,,
\end{equation}
with the coefficient probably good to about 20 \%.

We also test the sensitivity of the results to the end point: carrying
out the integration all the way to $r_\mathrm{S}=2\,m$, as in our
second model, Eqs. (\ref{dEJdt})--(\ref{y0}), has only a one percent effect on
the velocity for $\eta = 0.2$, and has essentially negligible effect
for smaller
values of $\eta$. We also vary the value of the radius where we match
the adiabatic part of the velocity with the beginning of the plunge
integration. For matching radii between $5.3\,m$ and $6\,m$, the final
kick velocity varies by at most seven percent for $\eta=0.2$ and five
percent for $\eta=0.1$.

In establishing the initial energy for the plunge orbit, we used the
quadrupole approximation for $dr_\mathrm{H}/dt$ in harmonic coordinates.
We have repeated the computation using a 2PN expression for
$dr_\mathrm{H}/dt$ expressed in terms of $m\,\omega$; the effect of the
change is negligible.

Our second method for matching to the plunge orbit,
Eqs. (\ref{dEJdt})--(\ref{y0}), gives virtually identical results. For
the 2PN correct values, and for values of the parameter $\alpha$ below
$0.1$, this method gives velocities that are in close agreement with
those shown in Figure \ref{fig1}.  For instance, with $\alpha=0.1$ and
$\eta=0.2$, the kick velocity is equal to $245 \, \mathrm{km \,
s}^{-1}$, compared to $243 \, \mathrm{km \, s}^{-1}$ with the first
method.  Small values of $\alpha$ correspond to a smoother match
between the circular orbit at the ISCO and the plunge orbit.  For
$\alpha=1$, implying a cruder match, the kick velocities are lower
than those shown in Figure \ref{fig1}: 4 \% lower for $\eta=0.1$, 10
\% lower for $\eta=0.2$, and 14 \% lower for $\eta=0.24$. These
differences are still within our overall error estimate of about 20 \%
indicated in Figure \ref{fig1}.

\section{Concluding remarks}
\label{secV}

Our results are consistent with, but substantially sharper than the
estimates for kick velocity for non-spinning binary black holes given by
\cite{Favata04}.  They are also consistent with estimates given by
\cite{manuela05} obtained from the Lazarus program for studying binary black
hole inspiral using a mixture of perturbation theory and numerical
relativity.  A recent improved analysis \citep{manuela05} gives 
$240 \pm 140 \, \mathrm{km \, s}^{-1}$ at $\eta = 0.22$ and
$190 \pm 100 \, \mathrm{km \, s}^{-1}$ at $\eta = 0.23$; as compared with our
estimates of $211 \pm 40$ and $183 \pm 37$, respectively.  In the limit of
small mass ratio, Eq. (\ref{pertlimit}) 
agrees very well with the result 
$V/c = 0.045 \eta^2$ obtained by \cite{NakaOoKoj87} using black hole
perturbation theory for a head-on collision from infinity.  
Since, as we have
seen, the contribution of the inspiral phase is small and the recoil is
dominated by the final plunge, one might expect
a calculation
of the recoil from a head-on plunge to be roughly consistent with
that from a plunge following an inspiral,
despite the different initial conditions; accordingly 
the agreement
we find with \cite{NakaOoKoj87} for the recoil values is 
satisfying\footnote{We thank Marc Favata
for bringing the \cite{NakaOoKoj87} work to our attention}.

Finally, we remark on the curious fact that our 2PN result shown in
Figure \ref{fig1} can be fit to better than one percent accuracy over
the entire range of $\eta$ by the simple formula
\begin{equation}\label{phenom}
\frac{V}{c} = 0.043 \,\eta^2 \sqrt{1-4\eta} \,
\left(1+\frac{\eta}{4}\right) \qquad \rm{(phenomenological)} \,.
\end{equation}
While we ascribe no special physical significance to this formula in
view of the uncertainties in our PN expansion, it illustrates that,
beyond the overall $\eta^2 \, \sqrt{1-4\eta}$ dependence, the
post-Newtonian corrections and the plunge orbit generate relatively
weak dependence on the mass ratio.  Such an analytic formula may be
useful in astrophysical modeling involving populations of binary black
hole systems.

Inclusion of the effects of spin will alter the result in several ways.
First, it will allow a net kick velocity even for equal mass black
holes. Second, it will significantly change the plunge orbits, depending
on whether the smaller particle orbits the rotating black hole in a
prograde or retrograde sense. In future work, we plan to treat this
problem using our 2PN formulae for linear momentum flux, augmented by
the 1.5PN spin orbit flux terms of \cite{kidder}, combined with a
similar treatment of plunge orbits in the equatorial plane of the Kerr
geometry.

\acknowledgments We are grateful to Jeremiah P. Ostriker for motivating us to
look at this question and for useful discussions, to Masaru Shibata
for useful input related to black hole perturbation calculations of
momentum recoil, and to Scott Hughes and Marc Favata for useful comments on
the manuscript. This work is supported in part by the US National
Science Foundation under grant No. PHY 03-53180. CW is grateful for the
support of the Centre National de la Recherche Scientifique, and the
hospitality of the Institut d'Astrophysique de Paris, where this work
was completed. MSSQ thanks B. R. Iyer and K. G. Arun for useful
discussions.  The calculations were performed with the help of the
softwares Mathematica and Maple.

\bibliography{BQW05}

\end{document}